 \documentclass[12pt,draftcls,onecolumn]{IEEEtran}

 \usepackage{color}

\usepackage{graphicx}
\usepackage{amsfonts,amsmath,amssymb}
\usepackage{mathrsfs}

\newtheorem{theorem}{Theorem}
\newtheorem{lemma}{Lemma}

\newtheorem{assumption}{Assumption}

\IEEEoverridecommandlockouts      

\usepackage{cite}

\title{\LARGE \bf
Guaranteed Cost LQG Control of Uncertain Linear
  Quantum Stochastic Systems\thanks{This work was
 supported by the
  Australian Research Council.
  A preliminary version of this paper has appeared in
  Proceedings of 2007 American Control Conference.}}

\author{A. J. ~Shaiju,
I. R.~Petersen\thanks{A.J. Shaiju (corresponding author) and I.R.
Petersen are with the School of Information Technology and
Electrical Engineering,
  University of New South Wales at the Australian Defence Force Academy,
    Canberra ACT 2600, Australia.
    Fax: +61-2-62688443,
     {\tt\small s.ainikkal@adfa.edu.au, i.petersen@adfa.edu.au.}}
 and
 M. R.~James\thanks{M.R. James is with the Department of
    Engineering, Australian
    National University, Canberra, ACT 0200,
    Australia.
  Fax: +61-2-61250506,
     Matthew.James@anu.edu.au} }
\begin{document}

\maketitle \thispagestyle{empty} 

\begin{abstract}
 In this paper, we formulate and solve a guaranteed cost control
problem for a class of uncertain linear stochastic quantum systems. For these
quantum systems, a connection with an associated classical
(non-quantum) system is first established. Using this connection, the desired
guaranteed cost results are established.
 The theory presented is illustrated using an example from quantum
 optics.
\end{abstract}


\section{INTRODUCTION}\label{sec:intro}
 The feedback control of quantum systems is  an important emerging
 research area; e.g., see \cite{AASDM-02, Bel-83,  
DJ-99,GSM-04,NJP1a, Wis-94,YK-03,YK-03a,BR-02,HJ-06,ZR-94}.
 However, most
of the existing results in quantum feedback control do not directly address
the issue of robustness with respect to parameter uncertainties in the
quantum system model. In this
paper, we study guaranteed cost control for a class of uncertain
linear quantum systems.
We consider  quantum systems described by linear
Heisenberg dynamics driven by quantum Gaussian noise processes, and
controlled by a  classical linear feedback controller. This class of
systems includes examples from quantum optics with classical
controllers implemented by standard analog or digital electronics.
For such quantum systems, we address the issue of robust controller
design by allowing for norm bounded uncertainties in the matrices
defining the quantum model. Also, we consider the case of uncertainties
in a quadratic Hamiltonian defining the quantum dynamics. Such uncertainties can represent
uncertainty in the values of the physical parameters of the quantum
system. Our results can also be extended to allow for linear
unmodeled quantum dynamics subject to a certain $H^\infty$ norm
bound constraint.

 Guaranteed cost control involves
constructing a controller such that the expected value of a
quadratic cost functional satisfies a given bound for all possible
values of the uncertain parameters in the model. The quadratic cost
functional is determined to reflect the performance requirements of
the quantum control system. This means that a controller can be
constructed which addresses not only the issue of robustness but
also the issue of performance.

A  classical
controller for a quantum systems takes  measurement data obtained by monitoring the
quantum system continuously in time. The controller determines the control actions which influence the
dynamics of the quantum system in a feedback loop.  The results in
this paper provide a method for designing such classical controllers  to
achieve a guaranteed bound on a quadratic  cost functional  when the
quantum  system model is subject to uncertainty. Our results exploit simple
computations of quantum expectations in Gaussian states which provide
a link to an auxiliary classical system. This enables us
to use established classical guaranteed cost control  results to obtain corresponding
quantum guaranteed cost control results.  

  The   paper is organized as follows. Section
\ref{sec:prelim} sets up a basic framework for describing
the class of uncertain linear quantum systems under consideration.
 A quantum linear system  is then related to
an auxiliary classical (non-quantum) system. 
In Section \ref{sec:main-result}, we  present  a guaranteed cost result
for the auxiliary classical  uncertain  system and then use this
to establish our main result which is   a guaranteed cost result
for the linear quantum uncertain  system. An example from quantum
optics is given in Section 
\ref{sec:eg} to illustrate our main results. Some conclusions are given in Section \ref{sec:conclusion}.


\section{PROBLEM FORMULATION AND PRELIMINARY RESULTS}
\label{sec:prelim}
\subsection{Uncertain Linear Quantum System}
We consider an uncertain  linear quantum   system described by the
following   non-commutative stochastic differential equations:
\begin{eqnarray} \label{QS}
dx &=& ([A+B_0\Delta C_0]x + [B_1+B_0 \Delta D_0]u) ~dt + B_0~ dv; \; x(0)=x_0,  \nonumber \\
 \mu &=& C_1 x + D_{12} u,\nonumber \\
  dy &=& [C_2+D_{20} \Delta C_0] x ~dt + [D_{22}+ D_{20} \Delta D_0]u dt + D_{20} ~dv; \; y(0)=0.
\end{eqnarray}
Here $v(t)$ vector of self-adjoint quantum noises
with Ito table:
\begin{eqnarray}
dv(t) dv^T(t) &=& F_v dt,
\label{ito}
\end{eqnarray}
where $F_{v}$ is a non-negative Hermitian matrix which satisfies  $\frac{1}{2} (F_v+F_v^T)=I$; e.g., see
\cite{Par-92,Bel-91}. Note that it is straightforward to extend the results of this paper to allow for more general matrices $F_{v}$; e.g., see \cite{JNP07}.
The noise processes can be
represented as operators on an appropriate Fock space; e.g., see
\cite{Par-92,Bel-91}.   
Also
 $x(t)$
is  vector of possibly non-commutative self-adjoint system variables
which are operators defined on an appropriate Hilbert space. The components of $x(t)$
represent physical properties of the system at time $t$ (using the Heisenberg
picture), such as position and momentum. The quantity $\mu(t)$ is also
a  vector of possibly non-commutative self-adjoint variables
corresponding to
physical observables defining the desired performance
objective. The classical quantities $ u(t)$ and $y(t)$
 represent control inputs and measurement outputs, respectively.
Their components are self adjoint and commute among themselves and at
different times. 
The matrices $A$, $B_0$,  $B_1$,  $C_0$, $C_1$, $C_2$, $D_0$, $D_{12}$, $D_{22}$, $D_{20}$ are known
constant real matrices of appropriate order
 and $\Delta$ is an uncertain norm
 bounded  real matrix satisfying
\begin{equation}
\Delta^{T} \Delta \leq I.
\label{Delta}
\end{equation}

We also consider an associated quadratic cost functional 
\begin{equation} \label{QPI}
J(u(\cdot))= \int_0^{t_f} \langle   {\mu}^{T}(t) \mu(t) \rangle ~dt.
\end{equation}
where the notation $\langle \cdot \rangle$ represents
expectation over all initial variables and noises.
The interval  $[0,t_f]$ is the fixed time horizon.
It is assumed that the initial condition  $x_0$ is Gaussian, with density operator
$\rho$. We let
 $\check{x}_0 := \langle x_0 \rangle$ and
\begin{eqnarray*}
 Y_0 &:=&  \frac{1}{2} \langle  (x_0 -\check{x}_0 )(x_0-\check{x}_0)^T +
 ((x_0 -\check{x}_0 )(x_0-\check{x}_0)^T)^T   \rangle.
\end{eqnarray*}
 The system model (\ref{QS}) includes a wide
range of quantum, classical, and quantum-classical stochastic
uncertain systems.

Together with the uncertain quantum system (\ref{QS}), consider a classical
(non-quantum) controller:
\begin{eqnarray} \label{CC}
dx_K &=& A_Kx_K  ~dt + B_K~ dy; \; x_K(0)=x_{K0},~~  
 u = C_K x_K.
\end{eqnarray}
Here the controller initial state $x_{K0}$ is a fixed real vector.
For a given value of the uncertainty matrix $\Delta$ satisfying
(\ref{Delta}), the quantum system (\ref{QS}) and the
classical controller (\ref{CC}) together produce a closed loop
classical-quantum 
system:
\begin{eqnarray} \label{CLS}
d\eta &=& \tilde{A} \eta  ~dt +  \tilde{B}~ dv;
 \; \eta(0)=\eta_0,
\nonumber  \\
  \mu &=&
\tilde{C} \eta,
\end{eqnarray}
where
\begin{eqnarray}
\label{CL_matrices}
 \tilde{A}&=&
\left[
\begin{array}{ll}
A +B_0 \Delta C_0& [B_1+B_0 \Delta D_0]C_K \\ 
B_K[C_2+D_{20} \Delta C_0] & 
A_K+B_K[D_{22}+ D_{20} \Delta D_0]C_K
\end{array}
\right], \, \tilde{B}= \left[
\begin{array}{c}
B_0 \\ B_K D_{20}
\end{array}
\right], \, \nonumber \\
 \tilde{C}&=& \left[
\begin{array}{ll}
C_1 & D_{12}C_K
\end{array}
\right], \, \eta= \left[
\begin{array}{c}
x \\  x_K
\end{array}
\right].
\end{eqnarray}
The guaranteed cost control problem under consideration involves
constructing a classical output feedback controller of the form
(\ref{CC}) and a cost bound $M>0$, such that cost (\ref{QPI})
corresponding to the closed loop system (\ref{CLS}) satisfies the
bound 
$
J(u(\cdot)) \leq M
$
for all uncertainty matrices $\Delta$ satisfying (\ref{Delta}). 

For a given value of the uncertainty matrix $\Delta$ satisfying
(\ref{Delta}), we now
define the following matrix valued function of time associated with
the  closed loop system (\ref{CLS}):
\begin{equation} \label{Pdefn}
P(t):= \frac{1}{2}\langle  \eta(t) \eta^{T}(t) +   ( \eta(t)
\eta^{T}(t))^T \rangle.
\end{equation}
Note that $P_0:=P(0)=\mbox{diag}(Y_0,0)$.
 Using this definition, we can establish the following lemmas which
 provide a  link between the  cost 
associated 
with  the  classical-quantum closed loop system (\ref{CLS}) and the
cost associated with an  auxiliary linear classical (non-quantum)
closed loop system.  


\begin{lemma}  \label{lemma:cost-P}
 The value of the cost (\ref{QPI}) corresponding to the  closed loop
 system (\ref{CLS}) is given by 
\begin{equation} \label{cost}
J(u(\cdot))= \int_0^{t_f} \mbox{ Tr} ( \tilde{C}^T \tilde{C} P(t)
)~dt.
\end{equation}
\end{lemma}

{\em Proof.}
 We have
\[
\langle  \mu^{T} \mu  \rangle = \langle  \eta^{T} \tilde{C}^{T}
\tilde{C} \eta \rangle  =\langle  \mbox{Tr} (\eta^{T}
\tilde{C}^{T} \tilde{C} \eta) \rangle 
= \frac{1}{2}\langle  \mbox{Tr} (
\tilde{C}^{T} \tilde{C} [\eta \eta^{T} +(\eta \eta^{T})^T]) \rangle
=  \mbox{Tr} ( \tilde{C}^{T} \tilde{C} P).
\]
Hence, upon integration, it follows that the corresponding closed loop
  cost is given by  (\ref{cost}).
$\Box$



\begin{lemma}  \label{lemma:cost-Riccati}
The matrix valued function $P(\cdot)$ defined by (\ref{Pdefn})
satisfies the differential  equation
\begin{equation} \label{DRE}
\dot{P}= \tilde{A}P+ P \tilde{A}^{T}+ \tilde{B} \tilde{B}^{T}; \;
P(0)=P_0,
\end{equation}
where
$P_0=\mbox{diag}(Y_0,0)$.
\end{lemma}
{\em Proof.}
 Using the quantum Ito rule (e.g., see Chapter III of
\cite{Par-92}), it follows from  the definition of $P(\cdot)$ in
(\ref{Pdefn}), that
\begin{eqnarray*}
dP &=& \frac{1}{2} [ \langle  d\eta \, \eta^{T}  \rangle +
\langle (d\eta \, \eta^{T})^T  \rangle 
  + \langle \eta
\, d\eta^{T}  \rangle +  \langle (\eta \, d\eta^{T})^T
\rangle    
  + (\tilde{B} F_{v} \tilde{B}^{T} + (\tilde{B} F_{v}
\tilde{B}^{T})^T )~dt] \\
&=& [  \tilde{A} P  + P\tilde{A}^{T} +  \frac{1}{2}\tilde{B}
(F_{v} +F_v^T) \tilde{B}^{T}]~dt.
\end{eqnarray*}
 Therefore, $P(\cdot)$ satisfies the differential
equation (\ref{DRE}).
$\Box$

\subsection{Auxiliary Classical Uncertain System}
We now define the auxiliary classical uncertain system which will be
used to solve the quantum guaranteed cost control problem defined
above: 
\begin{eqnarray} \label{CS}
dx &=& ([A+B_0 \Delta C_0]x + [B_1+B_0 \Delta D_0]u)
~dt + B_0~ dv; \; x(0)=x_0, \nonumber  \\
 \mu &=& C_1 x + D_{12} u, \nonumber \\
  dy &=& [C_2+D_{20} \Delta C_0] x ~dt + [D_{22}+ D_{20} \Delta D_0]u dt + D_{20} ~dv; \; y(0)=0.
\end{eqnarray}
Here, $x(t)$ is a real vector of state variables, $u(t)$ is the control
input vector, $y(t)$ is the measured output vector, $w(t)$ is a
classical Wiener process, and $x_0$ is a Gaussian
random variable with mean $\check{x}_0$ and covariance matrix
$Y_0$. Also,  $\Delta$ is an uncertain norm
 bounded  real matrix satisfying (\ref{Delta}). 
The matrices defining the classical uncertain system (\ref{CS})
are the same as the matrices defining the quantum uncertain system
(\ref{QS}). Associated with this classical uncertain system is the
quadratic cost functional
$
\hat{J}(u(\cdot)):= \int_0^{t_f}
 {\mathbf{E}} [ \mu^T(t) \mu(t)]  ~dt.
$

For a given value of the uncertainty matrix $\Delta$ satisfying
(\ref{Delta}), the classical system (\ref{CS}) and the
classical controller (\ref{CC}) together produce a closed loop
classical system:
\begin{eqnarray} \label{CCLS}
d\zeta &=& \tilde{A} \zeta  ~dt +  \tilde{B}~ dv;
 \; \eta(0)=\eta_0,
\nonumber  \\
  \mu &=&
\tilde{C} \zeta,
\end{eqnarray}
where $\zeta :=
  \left[
  \begin{array}{c}
    x \\
    x_K \\
  \end{array}
\right]$ is a real vector of state variables for the closed
loop system and the matrices $\tilde{A}$,  $\tilde{B}$, and
$\tilde{C}$ are defined as in (\ref{CL_matrices}). 

\begin{lemma}  \label{lemma:cost-equivalence}
For a given  uncertainty matrix $\Delta$ satisfying
(\ref{Delta}), the value of the cost $J(u(\cdot))$ corresponding to
the quantum classical closed loop system (\ref{CLS}) is the same as
value of the cost $\hat{J}(u(\cdot))$ corresponding to the classical
closed loop system (\ref{CCLS}). 
 \end{lemma}

{\em Proof.}
Let
$
Q(t):=
   {\mathbf{E}} \zeta(t) \zeta^T(t). $ Clearly $Q(0)=P_0$.
 As in Lemma \ref{lemma:cost-P}, one can now show that
$
\hat{J}(u(\cdot))=\int_0^{t_f} \mbox{ Tr } ( \tilde{C}^T  \tilde{C}
Q(t))~dt.
$
Also, in a similar fashion to the proof of  Lemma \ref{lemma:cost-Riccati} (but using
the classical Ito rule instead of the quantum Ito rule), it follows that
$Q(\cdot)$ also satisfies the differential  equation (\ref{DRE}).
Thus $ Q(\cdot)\equiv P(\cdot), $
 and hence
$ \hat{J}(u(\cdot))=J(u(\cdot)) $.
$\Box$

From this lemma, we immediately obtain the following result. 

\begin{lemma}  \label{lemma:GCC-equivalence}
Suppose a classical controller of the form (\ref{CC}) is a guaranteed cost
controller for the classical uncertain system (\ref{CS}) such that the
closed loop system (\ref{CCLS}) satisfies $\hat{J}(u(\cdot)) \leq M$
for all uncertain matrices $\Delta$ satisfying (\ref{Delta}). Then
this classical controller is also  a guaranteed cost
controller for the quantum uncertain system (\ref{QS}) such that the
closed loop system (\ref{CLS}) satisfies ${J}(u(\cdot)) \leq M$
for all uncertain matrices $\Delta$ satisfying (\ref{Delta}).
\end{lemma}

\section{THE MAIN RESULT}
\label{sec:main-result}
It follows from the results of the previous section that we can find a
guaranteed cost controller for the quantum uncertain system (\ref{QS})
by constructing a guaranteed cost controller for the classical
uncertain system (\ref{CS}). This leads to the main result of this paper which is Theorem \ref{thm:2} given in Subsection 
\ref{subsec:guar_cost_quant}. However, to obtain this result, we first consider the classical case. 
\subsection{Guaranteed Cost Control of the Classical Uncertain System}
\label{subsec:guar_cost_class}
In order to construct a suitable guaranteed cost controller for the
classical system (\ref{CS}), we will use a result which is
derived from the minimax LQG results of \cite{PUS-00}. In order to
present this result, we first require some assumptions and notation. 

\begin{assumption}  \label{ass:A1}
 For simplicity, we assume
$
C_1= \left[ \begin{array}{c}  R^{1/2} \\ 0
\end{array} \right]$, and $D_{12}= \left[ \begin{array}{c}  0 \\
 G^{1/2}
\end{array} \right].
$
With this simplification, the expression for the cost becomes
\begin{equation}\label{QC}
\hat J(u(\cdot))=\int_0^{t_f} [x^{T} Rx+u^{T}Gu]~dt.
\end{equation}
Also, we assume there exists a  $d_0>0$ such that $ \Gamma  =
D_{20}D_{20}^T \geq d_0I$.
\end{assumption}

{\em Notation.}
 For $\tau >0$, we define the matrices
\[
 R_\tau = R + \tau C_0^TC_0, \; \;
G_\tau = G + \tau D_{0}^T D_{0},
 {\Upsilon}_\tau =  \tau C_0^TD_{0}.
\]

\begin{assumption}  \label{ass:A2}
There exists a $\tau >0$ such that the following three conditions
hold: 
\begin{enumerate}
\item The Riccati differential equation
\begin{eqnarray}
\label{rde1}
\dot{Y} &=& (A - B_0D_{20}^{T}\Gamma^{-1}C_{2})Y 
+
Y(A-B_0D_{20}^{T}\Gamma^{-1}C_{2})^{T} 
 -Y(C_{2}^{T}\Gamma^{-1}C_{2}-\frac{1}{\tau}  R_\tau)Y \nonumber \\
&&+
 B_0 (I-D_{20}^{T}\Gamma^{-1}D_{20}) B_0^{T},
\end{eqnarray}
has a symmetric solution $Y(\cdot): [0,t_f] \to {\mathbf{R}}^{n
\times n}$ satisfying
 $Y(0)=Y_0$ and,
 there exists a  $c_0 >0$, such that 
$
Y(t) \geq c_0I, \,\, \,\, \mbox{ for all } 0 \leq t \leq t_f.
$
\item
The Riccati differential equation
\begin{eqnarray}
\label{rde2}
- \dot{X} &=& X  (A - B_1 {G}^{-1}_\tau  {\Upsilon}_\tau^T  ) +
 (A - B_1 {G}^{-1}_\tau  {\Upsilon}_\tau^T  )^T X +
 R_\tau-{\Upsilon}_\tau{G}_\tau^{-1} \Upsilon_\tau^T  \nonumber \\
   && -X( B_{2} G_\tau^{-1}
B_{2}^{T} - \frac{1}{\tau} B_0B_0^{T})X,
\end{eqnarray}
has a symmetric nonnegative definite solution $X(\cdot): [0,t_f] \to
{\mathbf{R}}^{n \times n}$ with $X(t_f)=0$.

\item For every $0 \leq t \leq t_f$, the spectral radius
of the matrix $Y(t)X(t)$ is less than $\tau$.
\end{enumerate}
\end{assumption}

\begin{theorem} \label{thm:1}
 Suppose that the classical uncertain system (\ref{CS}) is such that Assumptions \ref{ass:A1} and \ref{ass:A2} are satisfied.
Then the  controller (\ref{CC}) defined by the matrices
\begin{eqnarray*}
A_K  &=& A  + \frac{1}{\tau} YR_\tau  - B_K C_{2}
   +  ( B_{1} + \frac{1}{\tau}Y{\Upsilon}_\tau ) C_K-B_KD_{22}C_K,
   \nonumber \\
B_K &=& (Y C_{2}^{T} + B_0 D_{20}^{T}) \Gamma^{-1},~~
C_K = -G_\tau^{-1} (B_{1}^{T} X  +
\Upsilon^T_\tau)(I-\frac{1}{\tau}YX )^{-1},
\end{eqnarray*}
and the initial condition $x_{K0} = \check{x}_0$, is a guaranteed cost controller for the  classical uncertain system
(\ref{CS}), and  
   the associated closed loop value of the cost satisfies the bound
 $\hat J(u(\cdot))
 \leq V_{\tau} $ for all uncertainty matrices $\Delta$  satisfying
 (\ref{Delta}) where 
\begin{eqnarray}
  2 V_{\tau}
 & = &
    \check{x}_{0}^{T} X(0) (I-\frac{1}{\tau}
Y_0X(0))^{-1} \check{x}_{0}
  \nonumber  \\
&& +
     \int_0^{t_f} \mbox{ Tr } \Big[
   YR_\tau 
+
 B_K  (Y C_{2}^{T} + B_0  D_{20}^{T}))^{T} X
(I-\frac{1}{\tau}YX)^{-1} \Big]~dt.
\end{eqnarray}
\end{theorem}

{\em Proof.}
First note  the classical uncertain system (\ref{CS}), can be
re-written in the form
\begin{eqnarray} \label{CUS}
dx &=& (Ax + B_{1}u + B_0 \xi) ~dt + B_0~ dw, \nonumber  \\
  \mu &=& C_1 x + D_{12} u, \nonumber \\
  z &=& C_0x+D_0u, \nonumber \\
    dy &=& (C_{2} x + D_{22}u + D_{20} \xi) ~dt + D_{20} ~dw,
\end{eqnarray}
where 
$
\xi = \Delta z
$.
Also, $ \| \xi \| = \| \Delta z \| \leq \| z \| $.
 This yields
the Stochastic Integral Quadratic Constraint:
\begin{equation} \label{IQC}
{\bf E} \int_0^{t_f} \| \xi(t) \|^2 ~dt \leq
 {\bf E} \int_0^{t_f} \|
z(t) \|^2 ~dt.
\end{equation}
It now follows that the uncertainty in the classical uncertain system
(\ref{CS}) is a  special case of the uncertainty in  the minimax optimal  control result Theorem
8.4.1 of \cite{PUS-00}. From this result the required guaranteed cost
control result follows.
$\Box$

\subsection{Guaranteed Cost Control of the Quantum Uncertain System}
\label{subsec:guar_cost_quant}
Combining Theorem \ref{thm:1} and Lemma \ref{lemma:GCC-equivalence},
we immediately obtain the following result which is the main result of
the paper. 

\begin{theorem}  \label{thm:2}
 Suppose that the quantum uncertain system (\ref{QS}) is such that Assumptions \ref{ass:A1} and \ref{ass:A2} are satisfied.
Then the  controller (\ref{CC}) defined by the matrices
\begin{eqnarray*}
A_K  &=& A  + \frac{1}{\tau} YR_\tau  - B_K C_{2}
   +  ( B_{1} + \frac{1}{\tau}Y{\Upsilon}_\tau ) C_K-B_KD_{22}C_K,
   \nonumber \\
B_K &=& (Y C_{2}^{T} + B_0 D_{20}^{T}) \Gamma^{-1},~~
C_K = -G_\tau^{-1} (B_{1}^{T} X  +
\Upsilon^T_\tau)(I-\frac{1}{\tau}YX )^{-1},
\end{eqnarray*}
and the initial condition $x_{K0} = \check{x}_0$, is a guaranteed cost controller for the  quantum uncertain system
(\ref{CS}), and the 
    associated closed loop value of the cost satisfies the bound
 $ J(u(\cdot))
 \leq V_{\tau} $ for all uncertainty matrices $\Delta$  satisfying
 (\ref{Delta}) where 
\begin{eqnarray}
\label{Bound}
  2 V_{\tau}
 & = &
     \check{x}_{0}^{T} X(0) (I-\frac{1}{\tau}
Y_0X(0))^{-1} \check{x}_{0}
  \nonumber  \\
&& +
     \int_0^{t_f} \mbox{ Tr } \Big[
   YR_\tau 
+
 B_K  (Y C_{2}^{T} + B_0  D_{20}^{T}))^{T} X
(I-\frac{1}{\tau}YX)^{-1} \Big]~dt.
\end{eqnarray}
\end{theorem}

{\em Remarks.}
 It is possible to extend  Theorem \ref{thm:1} to the
case in which the norm bounded uncertain matrix $\Delta$ in the
classical uncertain system (\ref{CS}) is
replaced by a stable $H^\infty$ norm bounded transfer function matrix
$\Delta(s)$; e.g., see Section 2.4.3 of \cite{PUS-00}. This enables
Theorem \ref{thm:2} to be extended to allow for linear unmodeled
dynamics in the quantum system (\ref{QS}). In this case, the linear
unmodeled quantum dynamics would be defined by quantum stochastic
differential equations corresponding to matrices obtained from a state
space realization of the uncertain transfer function $\Delta(s)$.

Note that if the quantum system (\ref{QS}) has no uncertainty, 
$C_0=0$, $D_0 = 0$, and we let $\tau \rightarrow \infty$, then Theorem
\ref{thm:2} reduces to a result on quantum LQG control; e.g., see
\cite{DJ-99,EB05}.


\subsection{Application to Quantum Uncertain Systems with 
Uncertainty  in the Hamiltonian Matrix } \label{qho}
Rather than describing a linear quantum system in terms of a quantum
stochastic differential equation such as  in (\ref{QS}), a quantum
system can also be described in terms of a quadratic
Hamiltonian $H=x(0)^T R_0 x(0)$  and a coupling  operator $L= \Lambda
x(0)$; e.g., see \cite{EB05,JNP07}. Here $R_0$ is an $n\times n$ real
matrix and $\Lambda$ is an $N_w \times n$ complex matrix. 
Then as in \cite{EB05,JNP07}, the dynamics of the quantum system can
be described as follows:
\begin{eqnarray*}
 x_k(t) &=& U(t)^*x_k(0)U(t), ~~~~ k=1,2,\cdots,n,\\
 y_l(t) &=& U(t)^*w_l(t)U(t), ~~~~ l=1,2,\cdots,n_y,\\
 dU&=& \Big(
 -\imath H ~dt -\frac{1}{2}L^\dagger L~dt+
  [-L^\dagger,L^T] \Gamma~dw
   \Big)U,~
 U(0) =I.
\end{eqnarray*}
where the variables $x_k$ are the system variables, the variables
$y_l$ are the output variables and $U(t)$ is an adapted process of
unitary operators. Then as in \cite{JNP07}, the  nominal  matrices $A,B_0, B_1, C_2,D_{20}$  in (\ref{QS}) are given by
 \begin{eqnarray*}
 A &=& 2 \Theta \Big( R_0 + \Im (\Lambda^\dagger \Lambda)
   \Big),  \\
 \left[ \begin{array}{cc} B_0 & B_1 \end{array} \right]
  &=& 2 \imath \Theta [-\Lambda^\dagger  \Lambda^T]
 \Gamma,   \\
 C_2 &=& P_{N_y}^T
  \left[ \begin{array}{cc}
 \Sigma_{N_y} & 0_{N_y \times N_w} \\
 0_{N_y \times N_w} & \Sigma_{N_y}
     \end{array} \right]
     \left[ \begin{array}{c}
   \Lambda + \Lambda^\# \\ -\imath \Lambda +
   \imath \Lambda^\#
       \end{array} \right],\\
     \left[ \begin{array}{cc}  D_{20}& 0_{n_y \times
      n_u} \end{array} \right] &=&
 P_{N_y}^T
  \left[ \begin{array}{cc}
 \Sigma_{N_y} & 0_{N_y \times N_w} \\
 0_{N_y \times N_w} & \Sigma_{N_y}
     \end{array} \right] P_{N_w} =
      \left[
     \begin{array}{cc}
 I_{n_y \times n_y} & 0_{n_y \times (n_w -n_y)}
     \end{array}
     \right],
\end{eqnarray*}
 where $N_w=\frac{n_w}{2}$,
  $N_y=\frac{n_y}{2}$,
  $\Sigma_{N_y} =
   \left[
     \begin{array}{cc}
 I_{N_y \times N_y} & 0_{N_y \times (N_w -N_y)}
     \end{array}
     \right]   $, $\Theta = \frac{1}{2\imath}
     \Big( x(0)x(0)^T -(x(0)x(0)^T)^T  \Big)$,
      $P_{N_y}$ is the $n_y \times n_y$ permutation matrix
       satisfying $$P_{N_y}
       \left[ \begin{array}{cccc}
        a_1 & a_2 &\cdots & a_{n_y} \end{array} \right]^T
         = \left[ \begin{array}{cccccccc}
        a_1 & a_3 &\cdots & a_{n_y-1} &
         a_2 & a_4 &\cdots &a_{n_y} \end{array} \right]^T,$$
 $\Gamma=P_{N_w}.  \mbox{diag}_{N_w}\Big( \frac{1}{2}
 \left[ \begin{array}{cc} 1 & \imath \\
 1 & -\imath  \end{array}  \right]
  \Big)$, and $\Lambda^\#$ is obtained by taking the adjoint of each
   of the  components of $\Lambda$. 

We now
consider the case in which 
the matrix  defining the Hamiltonian is subject to uncertainty with
a specific structure and show that this leads to an uncertain linear
quantum system of the form (\ref{QS}). 
Suppose that  the quadratic
Hamiltonian is of the form $H=x(0)^T R x(0)$ where 
\begin{equation}
\label{R_uncertainty}
R=R_0 + \imath
  [-\Lambda^\dagger \Lambda^T] \Gamma_0 \Delta C_0, 
 \Delta = \left[ \begin{array}{c}
   0_{n_y \times n} \\
    \tilde \Delta 
    \end{array} \right], 
\end{equation}
 and $\tilde \Delta$ is a real $(n_w -n_u - n_y)\times n$ uncertain
 matrix satisfying 
 $\tilde \Delta \tilde \Delta^T \leq I$. Here
     $C_0$ is an arbitrary (but fixed)
    $n \times n$ matrix, and $\Gamma_0$ is the matrix
     consisting of the first $n_w - n_u$ columns
      of $\Gamma$. 
In this case, it is straightforward to verify that this leads to a
linear quantum uncertain system of the form (\ref{QS}) where the
matrices $A$, $B_0$,  $B_1$, $C_0$, $C_2$, $D_{20}$ are defined
as above.


\section{ILLUSTRATIVE EXAMPLE}
\label{sec:eg}
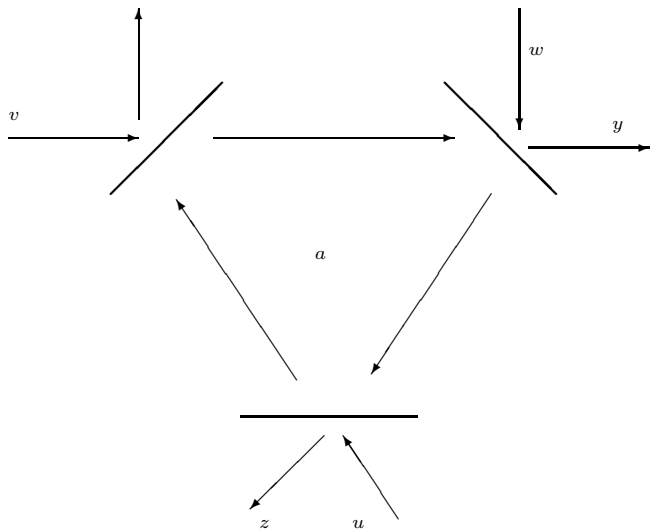
\begin{figure}[h]
\begin{center}
\setlength{\unitlength}{3068sp}%
\begingroup\makeatletter\ifx\SetFigFont\undefined%
\gdef\SetFigFont#1#2#3#4#5{%
  \reset@font\fontsize{#1}{#2pt}%
  \fontfamily{#3}\fontseries{#4}\fontshape{#5}%
  \selectfont}%
\fi\endgroup%
\begin{picture}(5199,4261)(1564,-4910)
\put(3601,-4861){\makebox(0,0)[lb]{\smash{{\SetFigFont{7}{8.4}{\familydefault}{\mddefault}{\updefault}{\color[rgb]{0,0,0}$z$}%
}}}}
\thicklines
{\color[rgb]{0,0,0}\put(5101,-1261){\line( 1,-1){900}}
}%
\thinlines
{\color[rgb]{0,0,0}\put(3226,-1711){\vector( 1, 0){1950}}
}%
{\color[rgb]{0,0,0}\put(5476,-2161){\vector(-2,-3){969.231}}
}%
{\color[rgb]{0,0,0}\put(3901,-3661){\vector(-2, 3){969.231}}
}%
\thicklines
{\color[rgb]{0,0,0}\put(3451,-3961){\line( 1, 0){1425}}
}%
\thinlines
{\color[rgb]{0,0,0}\put(1576,-1711){\vector( 1, 0){1050}}
}%
{\color[rgb]{0,0,0}\put(2626,-1561){\vector( 0, 1){900}}
}%
{\color[rgb]{0,0,0}\put(4126,-4111){\vector(-1,-1){600}}
}%
{\color[rgb]{0,0,0}\put(4726,-4786){\vector(-2, 3){450}}
}%
{\color[rgb]{0,0,0}\put(5776,-1786){\vector( 1, 0){975}}
}%
{\color[rgb]{0,0,0}\put(5701,-661){\vector( 0,-1){975}}
}%
\put(1576,-1561){\makebox(0,0)[lb]{\smash{{\SetFigFont{7}{8.4}{\familydefault}{\mddefault}{\updefault}{\color[rgb]{0,0,0}$v$}%
}}}}
\put(5776,-1036){\makebox(0,0)[lb]{\smash{{\SetFigFont{7}{8.4}{\familydefault}{\mddefault}{\updefault}{\color[rgb]{0,0,0}$w$}%
}}}}
\put(4051,-2686){\makebox(0,0)[lb]{\smash{{\SetFigFont{7}{8.4}{\familydefault}{\mddefault}{\updefault}{\color[rgb]{0,0,0}$a$}%
}}}}
\put(6451,-1636){\makebox(0,0)[lb]{\smash{{\SetFigFont{7}{8.4}{\familydefault}{\mddefault}{\updefault}{\color[rgb]{0,0,0}$y$}%
}}}}
\put(4351,-4861){\makebox(0,0)[lb]{\smash{{\SetFigFont{7}{8.4}{\familydefault}{\mddefault}{\updefault}{\color[rgb]{0,0,0}$u$}%
}}}}
\thicklines
{\color[rgb]{0,0,0}\put(2401,-2161){\line( 1, 1){900}}
}%
\end{picture}%

\caption{\small An optical cavity (plant).} \label{fig:cavity1}
\end{center}
\end{figure}

  In this section, we consider  an example which
illustrates the use of
 Theorem \ref{thm:2}.
 We consider an optical cavity resonantly
coupled to three optical channels as shown
 in Figure \ref{fig:cavity1}; e.g., see \cite{GZ00} and \cite{JNP07}. 

The annihilation operator $a$ for this cavity system
 (representing a
 standing wave) evolves in time according to the equations
\begin{eqnarray} \label{cavity}
 da &=& -\frac{\gamma}{2}a~dt-\sqrt{\kappa_1}dW
 - \sqrt{\kappa_2}dV-\sqrt{\kappa_3}dU, \nonumber \\
 dY &=& \sqrt{\kappa_1} a ~dt +dW.
\end{eqnarray}
Here $\gamma=\kappa_1 + \kappa_2 + \kappa_3$.
The
system (\ref{cavity}) can be written in real quadrature form as follows
(e.g., see \cite{JNP07}):
\begin{eqnarray} \label{cavity-stdform}
dx &=& A x~dt + B_{1}u~dt  + B_0~ \left[
   \begin{array}{c}
     dw \\
     dv
   \end{array}
 \right], \nonumber \\
 dy &=& C_{2}x~dt + D_{20}~
 \left[
   \begin{array}{c}
     dw \\
     dv
   \end{array}
 \right].
\end{eqnarray}
Here $a= (x_1+ \imath x_2)/2$,
 $y=y_1 = Y+Y^*$,
 $V=(v_1+\imath v_2)/2$,
$W=(w_1+\imath w_2)/2$, $U=(u_1+\imath u_2)/2$,
\begin{eqnarray*}
du &=& udt =
\left[\begin{array}{l}u_1\\   u_2\end{array}\right]dt,~~
dv = \left[\begin{array}{l}dv_1\\  dv_2\end{array}\right],
\;
 dw =
\left[\begin{array}{l}dw_1 \\  dw_2\end{array}\right],
\end{eqnarray*}
\[
A=-\frac{\gamma}{2}I, \;
  B_0=-
  \left[
                              \begin{array}{cc}
                                \sqrt{\kappa_1}I_2  &
                                 \sqrt{\kappa_2}I_2
                              \end{array}
                            \right],
\]
\[
B_{1}=-\sqrt{\kappa_3} I, \;
 C_{2}=\left[
                                       \begin{array}{cc}
                                         \sqrt{\kappa_1} & 0 \\
                                       \end{array}
                                     \right],
D_{20}=\left[
         \begin{array}{cccc}
           1 &
           0 &
           0 &
           0
         \end{array}
       \right]^T.
\]
The quantum noises $v,w$  have   Hermitian Ito matrices defined as follows:
\[
dv(t) dv^T(t) = F_v dt,~~ dw(t) dw^T(t) = F_w dt,~~F_v=F_w= \left[
           \begin{array}{cc}
             1 & i \\
             -i & 1 \\
           \end{array}
         \right].
\]

Now suppose an uncertain parameter $\delta$ is introduced into the
linear quantum system (\ref{cavity-stdform}) corresponding to a
perturbation in the parameter $\kappa_2$. 
Then linear quantum system (\ref{cavity-stdform}) becomes:
\begin{eqnarray} \label{cavity-perturbed}
dx &=& (A-\frac{ \delta}{2}I) x~dt + B_{1}u~dt  + B_0(\delta)~ \left[
   \begin{array}{c}
     dw \\
     dv
   \end{array}
 \right], \nonumber \\
 dy &=& C_{2}x~dt + D_{20}~
 \left[
   \begin{array}{c}
     dw \\
     dv
   \end{array}
 \right].
\end{eqnarray}
where $B_{0}(\delta):=-
  \left[
                              \begin{array}{cc}
                                \sqrt{\kappa_1 }I
                              & \sqrt{\kappa_2 + \delta}I
                              \end{array}
                            \right]$. 
We assume that the absolute value of the uncertain parameter $\delta$
is bounded as $|\delta| \leq  \delta_0$ where
$ \delta_0 \leq
 2 \sqrt{1+\kappa_2}.
 $
Now let \[
\Delta= \frac{\delta}{2} \left[
  \begin{array}{c}
    0  \\ 
 \frac{1}{ \sqrt{\kappa_2 +\delta_0}} I
  \end{array}
\right], \;
 C_0=I, \; D_0=0
\]
and observe that $ B_0(\delta_0)\Delta C_0=-\frac{\delta}{2}I$, 
$D_{20}\Delta C_0 =0$,  $B_0(\delta_0)\Delta D_0=0$,  $D_{20}\Delta
D_0 = 0$, and  $\Delta^T \Delta \leq I$.
From this, it follows that the above model for the
  cavity is a special
 case of the following linear quantum uncertain system
\begin{eqnarray} \label{examplesys1}
dx &=& ([A+B_0(\delta_0)\Delta C_0]x + [B_{1}+B_0(\delta_0)\Delta D_0] u) ~dt + B_0(\delta) \left[
   \begin{array}{c}
     dw \\
     dv
   \end{array}
 \right], \nonumber  \\
    dy &=& ([C_{2}+D_{20}\Delta C_0]x+D_{20}\Delta
D_0u) ~dt + D_{20} \left[
   \begin{array}{c}
     dw \\
     dv
   \end{array}
 \right],
\end{eqnarray}
where  $\Delta^T \Delta \leq I$. Furthermore, in order to convert this
into a quantum uncertain system of the form (\ref{QS}) note that 
$
B_0(\delta) B_0(\delta)^T \leq 
 B_0(\delta_0) B_0(\delta_0)^T \mbox{ for all } \delta \mbox{
   such that } |\delta|
 \leq \delta_0.$
Hence, we
can increase the size of the noise in this uncertain system to obtain
the following quantum uncertain system of the form  (\ref{QS}):
\begin{eqnarray} \label{examplesys2}
dx &=& ([A+B_0(\delta_0)\Delta C_0]x + [B_{1}+B_0(\delta_0)\Delta D_0] u) ~dt + B_0(\delta_0) \left[
   \begin{array}{c}
     dw \\
     dv
   \end{array}
 \right], \nonumber  \\
    dy &=& ([C_{2} +D_{20}\Delta C_0]x+D_{20}\Delta
D_0u) ~dt + D_{20} \left[
   \begin{array}{c}
     dw \\
     dv
   \end{array}
 \right],
\end{eqnarray}
Thus, if we can construct a controller which leads to a guaranteed cost
upper bound on the cost functional
\begin{equation}\label{QC_ex}
J(u(\cdot))=\int_0^{t_f} \langle x^{T} Rx+u^{T}Gu\rangle ~dt,
\end{equation}
for this quantum uncertain system, then this  controller will lead
to the same upper bound on the closed loop value of the cost functional (\ref{QC_ex}) when applied to the
model of our cavity.

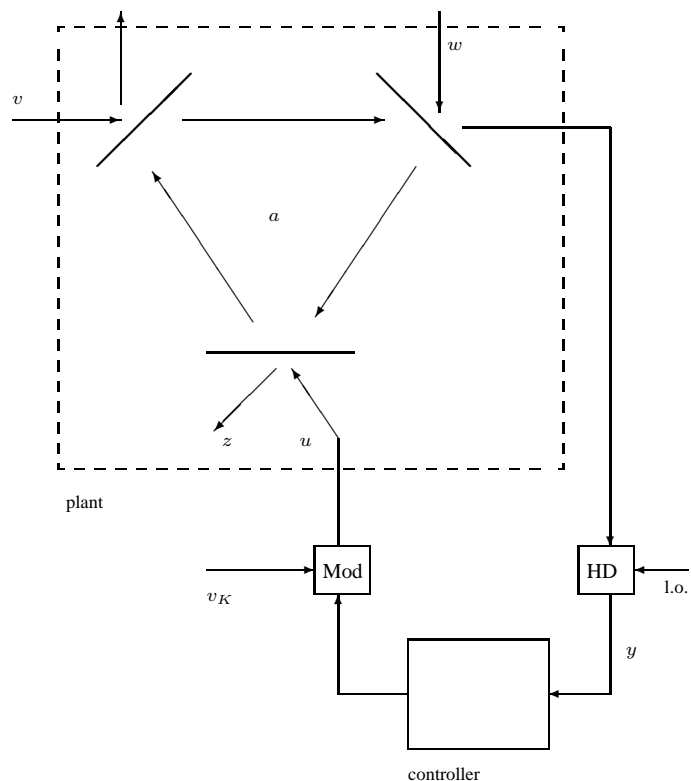
\begin{figure}[h]
\begin{center}
\setlength{\unitlength}{2568sp}%
\begingroup\makeatletter\ifx\SetFigFont\undefined%
\gdef\SetFigFont#1#2#3#4#5{%
  \reset@font\fontsize{#1}{#2pt}%
  \fontfamily{#3}\fontseries{#4}\fontshape{#5}%
  \selectfont}%
\fi\endgroup%
\begin{picture}(6549,7437)(1564,-8086)
\put(7876,-6286){\makebox(0,0)[lb]{\smash{{\SetFigFont{7}{8.4}{\familydefault}{\mddefault}{\updefault}{\color[rgb]{0,0,0}l.o.}%
}}}}
\thicklines
{\color[rgb]{0,0,0}\put(5101,-1261){\line( 1,-1){900}}
}%
\thinlines
{\color[rgb]{0,0,0}\put(3226,-1711){\vector( 1, 0){1950}}
}%
{\color[rgb]{0,0,0}\put(5476,-2161){\vector(-2,-3){969.231}}
}%
{\color[rgb]{0,0,0}\put(3901,-3661){\vector(-2, 3){969.231}}
}%
\thicklines
{\color[rgb]{0,0,0}\put(3451,-3961){\line( 1, 0){1425}}
}%
\thinlines
{\color[rgb]{0,0,0}\put(1576,-1711){\vector( 1, 0){1050}}
}%
{\color[rgb]{0,0,0}\put(2626,-1561){\vector( 0, 1){900}}
}%
{\color[rgb]{0,0,0}\put(4126,-4111){\vector(-1,-1){600}}
}%
{\color[rgb]{0,0,0}\put(4726,-4786){\vector(-2, 3){450}}
}%
{\color[rgb]{0,0,0}\put(5701,-661){\vector( 0,-1){975}}
}%
{\color[rgb]{0,0,0}\put(4726,-4786){\line( 0,-1){1050}}
}%
{\color[rgb]{0,0,0}\put(5926,-1786){\line( 1, 0){1425}}
}%
{\color[rgb]{0,0,0}\put(2026,-5086){\dashbox{100}(4875,4275){}}
}%
{\color[rgb]{0,0,0}\put(7351,-1786){\vector( 0,-1){4050}}
}%
{\color[rgb]{0,0,0}\put(7051,-6286){\framebox(525,450){}}
}%
{\color[rgb]{0,0,0}\put(4501,-6286){\framebox(525,450){}}
}%
{\color[rgb]{0,0,0}\put(7351,-6286){\line( 0,-1){975}}
\put(7351,-7261){\vector(-1, 0){600}}
}%
{\color[rgb]{0,0,0}\put(5401,-7261){\line(-1, 0){675}}
\put(4726,-7261){\vector( 0, 1){975}}
}%
{\color[rgb]{0,0,0}\put(5401,-7786){\framebox(1350,1050){}}
}%
{\color[rgb]{0,0,0}\put(3451,-6061){\vector( 1, 0){1050}}
}%
{\color[rgb]{0,0,0}\put(8101,-6061){\vector(-1, 0){525}}
}%
\put(1576,-1561){\makebox(0,0)[lb]{\smash{{\SetFigFont{7}{8.4}{\familydefault}{\mddefault}{\updefault}{\color[rgb]{0,0,0}$v$}%
}}}}
\put(5776,-1036){\makebox(0,0)[lb]{\smash{{\SetFigFont{7}{8.4}{\familydefault}{\mddefault}{\updefault}{\color[rgb]{0,0,0}$w$}%
}}}}
\put(4051,-2686){\makebox(0,0)[lb]{\smash{{\SetFigFont{7}{8.4}{\familydefault}{\mddefault}{\updefault}{\color[rgb]{0,0,0}$a$}%
}}}}
\put(4351,-4861){\makebox(0,0)[lb]{\smash{{\SetFigFont{7}{8.4}{\familydefault}{\mddefault}{\updefault}{\color[rgb]{0,0,0}$u$}%
}}}}
\put(3601,-4861){\makebox(0,0)[lb]{\smash{{\SetFigFont{7}{8.4}{\familydefault}{\mddefault}{\updefault}{\color[rgb]{0,0,0}$z$}%
}}}}
\put(2101,-5461){\makebox(0,0)[lb]{\smash{{\SetFigFont{7}{8.4}{\familydefault}{\mddefault}{\updefault}{\color[rgb]{0,0,0}plant}%
}}}}
\put(4576,-6136){\makebox(0,0)[lb]{\smash{{\SetFigFont{8}{9.6}{\familydefault}{\mddefault}{\updefault}{\color[rgb]{0,0,0}Mod}%
}}}}
\put(7126,-6136){\makebox(0,0)[lb]{\smash{{\SetFigFont{8}{9.6}{\familydefault}{\mddefault}{\updefault}{\color[rgb]{0,0,0}HD}%
}}}}
\put(5401,-8086){\makebox(0,0)[lb]{\smash{{\SetFigFont{7}{8.4}{\familydefault}{\mddefault}{\updefault}{\color[rgb]{0,0,0}controller}%
}}}}
\put(7501,-6886){\makebox(0,0)[lb]{\smash{{\SetFigFont{7}{8.4}{\familydefault}{\mddefault}{\updefault}{\color[rgb]{0,0,0}$y$}%
}}}}
\put(3451,-6361){\makebox(0,0)[lb]{\smash{{\SetFigFont{7}{8.4}{\familydefault}{\mddefault}{\updefault}{\color[rgb]{0,0,0}$v_{K}$}%
}}}}
\thicklines
{\color[rgb]{0,0,0}\put(2401,-2161){\line( 1, 1){900}}
}%
\end{picture}%

\caption{\small An optical cavity  (plant) controlled by a classical
system
  (controller $K$, implemented using standard electronics). The quadrature measurement
  is achieved by homodyne photo-detection (HD), and the control actions
  are applied via an optical modulator (Mod).}
\label{fig:cavity3}
\end{center}
\end{figure}

We now apply  Theorem \ref{thm:2} to the quantum uncertain system
(\ref{examplesys2}) and cost functional (\ref{QC_ex}), taking
$\kappa_1=\kappa_2=\kappa_3=2$, $R=G=I$, $t_f = 100$, and
$\delta_0=1$. For the long time horizon being considered, the
solutions to the Riccati differential equations (\ref{rde1}),
(\ref{rde2}) can be approximated by the solutions to the
corresponding algebraic Riccati equations. Solving these Riccati
equations for different values of the parameter $\tau
>0$, we find that the cost bound $V_\tau$ defined in (\ref{Bound})
is minimized with $\tau = 1.41$. For this value of $\tau$, we obtain
the cost bound of $V_\tau= 322.1$.  The corresponding Riccati
solutions are
\[
X= \left[\begin{array}{ll}
   0.455  & 0\\
         0 &   0.455
\end{array}\right];~~
Y= \left[\begin{array}{ll}
    1.267 &  0\\
         0  &  1.361
\end{array}\right].
\]
Also, the corresponding  controller matrices are
\begin{eqnarray*}
A_K&=&\left[\begin{array}{ll}
   -2.908   & 0\\
         0  & -2.297
\end{array}\right]; \;
B_K=\left[
                 \begin{array}{c}
    0.377 \\
         0
                 \end{array}
               \right],
C_K=\left[\begin{array}{ll}
    1.088   &0\\
         0   & 1.148
\end{array}\right].
\end{eqnarray*}
This classical controller  can be
implemented using standard electronic devices. The closed loop
quantum-classical system is illustrated in Figure
\ref{fig:cavity3}.

 We now consider a modification to the above  example to provide an
 example of a quantum system with uncertainty in the Hamiltonian
 matrix such as considered 
  in subsection \ref{qho}. This example consists of a cavity with
  uncertainty in the
  detuning parameter. The cavity detuning corresponds to  a 
  mismatch $\Omega$ between the resonant frequency of the optical
     cavity and the frequency of
    input field. In this case,  equation (\ref{cavity}) describing the
    cavity is 
    modified  by replacing the term
   $\frac{-\gamma}{2}$ by the term $\frac{-\gamma}{2}+
    \imath \Omega$.
 Thus, the 
  coefficient matrices are the same as in the previous case except the matrix 
   $A$ is now given by 
    $A = \left[ \begin{array}{cc}
 -\frac{\gamma}{2} & -\Omega \\
  \Omega & -\frac{\gamma}{2}
    \end{array}  \right]$.
 
This system can be considered as an open
 quantum
  harmonic oscillator with noise input $[w~ v~ u]^T$, 
 Hamiltonian matrix $R=\frac{-\Omega}{2}I$
  and  coupling matrix $\Lambda =
   \left[ \begin{array}{cc}
  \sqrt{\kappa_1} & \sqrt{\kappa_1} \imath \\
   \sqrt{\kappa_2} & \sqrt{\kappa_2} \imath \\
   \sqrt{\kappa_3} & \sqrt{\kappa_3} \imath
   \end{array}  \right]$. In this example, we consider an uncertainty in
 the frequency mismatch $\Omega$. Indeed,  a perturbation in $\Omega$
 to  $\Omega = \Omega_0 + \Omega_e$ with  
  $|\Omega_e| \leq \epsilon_0$  corresponds to a
perturbation     $R = R_0 +E$ in the Hamiltonian matrix
  where $R_0= \frac{-\Omega_0}{2}I$ and
  $E=\frac{-\Omega_e}{2}I$. Furthermore, this perturbation is of the
  form  (\ref{R_uncertainty})
  with $\tilde \Delta =
 \left[
   \begin{array}{cc}
     0 & -\frac{\Omega_e}{\epsilon_0} \\
 \frac{\Omega_e}{\epsilon_0} & 0 \\
   \end{array}
 \right]
   $
  and $C_0 = \frac{\epsilon_0}{\sqrt{\kappa_2}} I $.

The corresponding guaranteed cost controller was calculated in    the case when
 $\Omega_0=0$ and $\epsilon_0=1$  keeping all other system
  parameters the same as in the previous example. In this case, a parameter
  value of $\tau=0.9$ gave the
   minimal closed loop  cost bound $V_\tau=126$ and the associated controller
    matrices are
\begin{eqnarray*}
A_K&=&\left[\begin{array}{ll}
   -2.067   & 0\\
         0  & -2.336
\end{array}\right]; \;
B_K=\left[
                 \begin{array}{c}
    0.202 \\
         0
                 \end{array}
               \right],
C_K=\left[\begin{array}{ll}
    0.519   &0\\
         0   & 0.521
\end{array}\right].
\end{eqnarray*}
This controller could also be implemented as in Figure \ref{fig:cavity3}.

\section{CONCLUSIONS}
\label{sec:conclusion}

In this paper, we presented a theory for  synthesizing classical guaranteed cost
controllers for a class of uncertain linear quantum  stochastic
systems. The theory  was  illustrated using some simple examples from
quantum optics.



\begin{thebibliography}{IEEEtran}
\bibitem{AASDM-02}
M. A. Armen, K. J. Au, J. K. Stockton, A. C. Doherty and H. Mabuchi,
Adaptive homodyne measurement of optical phase,
 {\it Phy. Rev. A},
  89 (13), 2002.






\bibitem{Bel-83}
V. P. Belavkin, On the theory of controlling observable quantum
systems, {\it Automation and Remote Control}, 44(2): 178-188, 1983.



\bibitem{Bel-91}
V. P. Belavkin, continuous non-demolition observation, quantum
filtering and estimation, In {Quantum aspects of Optical
Communication}, Vol. 45, {\it Lecture Notes in Physics}, pp.
131-145, Springer, Berlin, 1991.


\bibitem{BHJ06a}
L. Bouten, R. Van Handel and M. R. James, An introduction to quantum
filtering, {arxiv.org/math.OC/0601741}, 2006.

\bibitem{BR-02}
E. Brown and H. Rabitz, Some mathematical and algorithmical
challenges in the control of quantum dynamics phenomena, {\it J.
Mathematical Chemistry},  31(1): 17-63, 2002.


\bibitem{HJ-06}
C. D'Helton and M. R. James, Stability, gain, and robustness in
quantum feedback networks, {\it Phy. Rev. A}, to appear.
quant-ph/0511140, 2006.


\bibitem{DJ-99}
A. C. Doherty and K. Jacobs, Feedback-control of quantum systems
using continuous state estimation, {\it Phy. Rev. A}, 60: 2700,
1999, quant-ph/9812004, 2006.

\bibitem{EB05}
S. C. Edwards and V. P. Belavkin,
 Optimal quantum  feedback control via
 quantum dynamic programming,
 quant-ph/0506018, University of Nottingham, 2005.

\bibitem{GZ00}
C. W. Gardiner and P. Zoller, {\it Quantum Noise}, Springer, Berlin,
2000.




\bibitem{GSM-04}
J. M. Geremia, J. K. Stockton and H. Mabuchi, Real-time quantum
feedback control of atomic spin-squeezing, {\it Science}, 304:
270-273, April 2004.



\bibitem{JNP07}
M. R. James, H. I. Nurdin and I. R. Petersen,
 $H^\infty$ Control of Linear Quantum Stochastic systems,
  to appear in {\it IEEE Transactions on Automatic Control}.

\bibitem{Llo-00}
S. Lloyd, Coherent quantum feedback, {\it Phy. Rev. A}, 62: 022108,
2000.

\bibitem{NJP1a}
H.~I. Nurdin, M.~R. James, and I.~R. Petersen.
\newblock Quantum {LQG} control with quantum mechanical controllers.
\newblock In {\em Proceedings of the 17th IFAC World Congress}, Seoul, Korea,
  July 2008.
\newblock To Appear, also see arXiv:0711.2551v1 [quant-ph].


\bibitem{Par-92}
K. R. Parthasarathy, {\it  An Introduction to Quantum Stochastic
Calculus}, Birkhauser, 1992.


\bibitem{PUS-00} I. R. Petersen, V. Ugrinovskii and A. V. Savkin, {\it
Robust Control Design Using $H^{\infty}$ Methods}, Springer, 2000.




\bibitem{Wis-94}
H. Wiseman, Quantum theory of continuous feedback, {\it Phy. Rev.
A}, 49(3): 2133-2150, 1994.


\bibitem{YK-03}
M. Yanagisawa and H. Kimura, Transfer function approach to quantum
control, Part I: Dynamics of quantum feedback systems, {\it IEE
Trans. Automatic Control}, 48(12):2107-2120, 2003.



\bibitem{YK-03a}
M. Yanagisawa and H. Kimura, Transfer function approach to quantum
control, Part II: Control Concepts and applications, {\it IEE Trans.
Automatic Control}, 48(12):2107-2121, 2132.




\bibitem{ZR-94}
H. Zhang and H. Rabitz, Robust optimal control of quantum molecular
systems in the presence of disturbances and uncertainties, {\it
Phys. Rev. A}, 49(4):2241-2254, 1994.



\end{thebibliography}
 \end{document}